\documentclass{jetpl}
\usepackage{hyperref}
\usepackage[numbers]{natbib}
\usepackage{textcomp}
\twocolumn

\renewcommand{\Re}{\mathrm{Re }}
\renewcommand{\Im}{\mathrm{Im }}
\renewcommand{\vec}{\mathbf }
\newcommand{\R}{\mathbb R}
\renewcommand{\C}{\mathbb C}

\newcommand{\E}{\mathrm e}
\newcommand{\diver}{\mathrm{div}}
\newcommand{\sign}{\mathrm{sign\,}}
\renewcommand{\D}{\mathrm d}
\newcommand{\I}{\mathrm i}
\newcommand{\pD}{\partial }

\newcommand{\red}{ }

\title{Fourth branch of instability of Stokes' wave and dependence of corresponding growth rate on nonlinearity}

\rtitle{Fourth branch of instability of Stokes' wave\ldots}

\sodtitle{Fourth branch of instability of Stokes' wave and dependence of corresponding growth rate on nonlinearity}

\author{A.\,O.~Korotkevich\/\thanks{e-mail: A.Korotkevich@Skoltech.ru},
A.\,O.~Prokofiev}

\rauthor{A.\,O.\,Korotkevich A.O., A.\,O.\,Prokofiev A.O.}

\sodauthor{Korotkevich, Prokofiev}

\address{Skolkovo Institute of Science and Technology, Bolshoy Boulevard 30, bld. 1, Moscow 121205, Russia\\~\\
and L.\,D.~Landau Institute for Theoretical Physics RAS,\\
 prosp. Akademika Semenova 1A, Chernogolovka, Moscow region, 142432, Russia}

\dates{20 November 2025}{9 December 2025}

\abstract{Through a massive computation we reached the fourth superharmonic instability branch of the Stokes' wave. Using the obtained results we checked phenomenological formulae for the dependence of the instability growth rates corresponding to different branches of instability on the nonlinearity parameter (steepness, defined as the wave \red{hight} to wavelength ratio $H/\Lambda$) in the vicinity of the new instability branch appearance and far from it. \red{It is demonstrated, that the formulae, obtained as a least squares fit (using the information from the first three branches of instability) and a phenomenological asymptotics, work for the fourth branch as well.} Range of applicability of the relations \red{is} corrected. \red{This result removes the necessity to compute further branches of instability if accuracy better than 10\% for the growth rate is acceptable.} Growth rates for all four instability branches are reported.}


\begin{document}
\maketitle

\textit{Introduction. } Investigation of Stokes' wave solution (1D periodic wave propagating with constant velocity without change of shape, capillarity is neglected with respect to gravity) is a classical problem of hydrodynamics of ideal incompressible fluid and, perhaps, the first nonlinear solution of hydrodynamic problem in general~\cite{Stokes1847,Stokes1880}. It was conjectured by Longuett-Higgins~\cite{LHS1978-1} that the Stokes' wave becomes unstable when we approach the limiting Stokes wave (wave of the maximum amplitude, with formation of singularity on the crest). Specifically, for superharmonic instability (namely, instability of a perturbation with a wavelength shorter than the period/wavelength $\Lambda$ of the Stokes' wave) new branch of instability appears with increase of steepness $s=H/\Lambda$, defined as a ratio of a maximum \red{hight} $H$ of a wave (difference in surface elevation between the deepest point \red{of} trough and the crest) to $\Lambda$, at every extremum of a Hamiltonian (for the first two branches of superharmonic instability see~\cite{Tanaka1983,Saffman1985,LHT1997}, the third one was confirmed in~\cite{KLSD2023}). Recently~\cite{PNAS2023} it was shown that both subharmonic and superharmonic instabilities become equally important when we approach the limiting Stokes' wave with steepness $s_{max}$.

In the recent paper~\cite{KLSD2023} the detailed investigation of the superharmonic instability was presented. It was possible to show \red{(we investigate instability of small perturbations in a linear approximation, thus looking for exponential growth $\sim \E^{\lambda t}$ at the initial stage of instability development)} that for small growth rates $\lambda < 1$ right after the passing of every threshold in steepness $s_n$ (here $n$ is the number of instability branch), corresponding to another extremum of a Hamiltonian, and appearance of a new instability branch, the universal dependence can be proposed. The data from the first three branches of superharmonic instability was used for a nonlinear least squares fit of the following form:
\begin{equation}
\label{small_gr_scaling}
\lambda^2 \approx (b_0 + b_1 x + b_2 x^2 + b_3 x^3)\ln(x), 
\end{equation}
where $\lambda$ is a growth rate of instability, $x = (s_{max} - s)/(s_{max} - s_n)$ is normalized steepness, and values of the fit constants are $b_0=-0.140023$, $b_1=0.0366936$, $b_2=-0.0129251$, and $b_3=0.00125835$. In the same paper~\cite{KLSD2023} the following estimation of steepness of the limiting Stokes' wave was provided \red{$s_{max} = 0.141063483980 \pm 10^{-12}$}, the most accurate at the time of publication. Steepnesses corresponding to thresholds of new superharmonic instability branches appearance (extrema of Hamiltonian) for the first four branches were reported as follows (Table 1 in~\cite{KLSD2023}):
$s_1 = 0.136603556$, $s_2 = 0.140796584$, $s_3 = 0.141049627$, and $s_4 = 0.141062741$ with an absolute error of the values less or equal than $10^{-7}$.
For large growth rates $\lambda > 1$ far beyond the instability thresholds the asymptotic scaling:
\begin{equation}
\label{large_gr_scaling}
\lambda^2 \sim 1/(s_{max} - s),
\end{equation}
was proposed. Because all these phenomenological dependences were obtained using the data from the first three branches of superharmonic instability, it is necessary to verify them using the data from the next (fourth) branch. It should be noted that relative difference in steepness between the threshold of the fourth branch appearance $s_4$ and the steepness of the limiting Stokes' wave $s_{max}$ is $(s_{max} - s_4)/s_{max} \simeq 5.27\times 10^{-6}$, less than $0.001\%$, so it is almost a limiting wave, which results in extreme requirements on resolution even for the inhomogeneous grid~\cite{LDS2017}.

In this Letter we report the first data from the fourth branch of superharmonic instability of the Stokes' wave. The growth rates were computed for two Stokes' waves beyond the instability threshold $s_4$, available from the collection of high precision Stokes' waves at~\cite{DLK2016,PadePolesList}. For the steepest wave the relative difference from the limiting Stokes' wave was less than $3.59\times 10^{-6}$. For both waves we confirm that both phenomenological dependences \eqref{small_gr_scaling} and \eqref{large_gr_scaling} give reasonable approximation for the growth rates.

\textit{Basic Equations. } Here we will provide the shortened theoretical description closely following the one given in~\cite{DLK2013,DLK2016} and~\cite{KLSD2023}. We consider 2D potential flow (velocity is given by $\vec v = \vec \nabla \Phi$, here $\Phi$ is a velocity potential) of ideal incompressible (thus $\diver \vec v =\Delta \Phi = 0$) fluid under action of a constant gravity acceleration $g$ directed antiparallel to $y$-axis (from top to bottom, toward $-\infty$), neglecting capillarity. Fluid occupies the half plane $-\infty < y < \eta(x,t)$, here $\eta(x,t)$ is the deviation of the fluid surface from the unperturbed state, which is a flat surface $y=0$. We consider periodic case $-\Lambda/2 < x < \Lambda/2$ (recall that $\Lambda$ is a period of the Stokes' wave). Instead of considering time-dependent domain occupied by the fluid it is convenient to introduce time-dependent conformal mapping $z(w,t)=x(u,v;t) + \I y(u,v;t)$ of a fixed domain (lower complex half-plane $\C^-$) of the new variable $w=u+\I v,$ $u,v \in \R$ into a time-dependent fluid domain in the physical complex plane $z=x+\I y$.  Because we are interested in $\Lambda$-periodic solutions in $x$, we consider one spatial period ($-\pi<u<\pi$ and $-\infty < v < 0$) in $w$-plane as well. It is natural to introduce Fourier series:
\begin{align} 
f_k=&\frac{1}{2\pi}\int\limits_{-\pi}^{\pi}
f(u)\exp\left (-\I ku\right )\D  u,\;\;\label{ffourier}\\
f(u)=&\sum\limits_{k=-\infty}^{+\infty} f_k\exp\left (\I
ku\right ).\nonumber
\end{align}
Then, one can start from Tanveer-Dyachenko equations (here and further bar $\bar f$ means complex conjugation):
\begin{align}
\frac{\pD R}{\pD t} &= \I \left(U R_u - R U_u \right), \label{Reqn4}\\
\qquad U &=\hat P^-(R\bar V+\bar RV), \quad B= \hat P^-(|V|^2), \label{UBdef4}\\
\frac{\pD V}{\pD t} &= \I \left[ U V_u - RB_u\right ]+ g(R-1), \label{Veqn4}
\end{align}
Here the variables ($\Psi(x,t)=\Phi(x,\eta(x,t), t)$ is the velocity potential evaluated at the surface of the fluid $\eta$) are:
\begin{equation}
R \equiv \frac{1}{z_u}\quad\mbox{and}\quad V \equiv \frac{\I(\Psi+\I\hat H\Psi)_u}{z_u},
\end{equation}
which were originally introduced by S.~Tanveer in~\cite{Tanveer1991} for the periodic boundary conditions (BCs) and later independently
obtained by A.\,I.~Dyachenko in~\cite{Dyachenko2001} for the infinite domain with decaying BCs. We also introduced Hilbert transformation $(\hat H f(u))_k=\I\,\sign(k)\,f_k$ and the projector operator of any function $f(u)$ defined by the Fourier series \eqref{ffourier} into the space of functions analytic in the lower half plane $w\in \C^-$, specifically $\hat P^-\equiv \frac{1}{2}(1+\I\hat H)$.

Stokes' wave is a solution of equations~\eqref{Reqn4}-\eqref{Veqn4} propagating with a constant speed $c$ without change of shape, thus both $R$ and $V$ are functions of  $u-ct$ only. To study linear stability of Stokes' wave, we consider a small
perturbation of such a solution  $R$, $V$ of
equations~\eqref{UBdef4}-\eqref{Veqn4} in the following form $R\rightarrow
R+\delta R$, $V\rightarrow V+\delta V$. After linearization of Eqs. \eqref{Reqn4}-\eqref{Veqn4} with respect to perturbations $\delta R$ and
$\delta V$, propagating with the wave, it is reasonable to look for exponential dependence of perturbations on time:
\begin{equation}\label{dRdVlambda}\begin{split}
& \delta R(u-ct,t)=e^{\lambda t}\delta R_{1}(u-ct)+e^{\bar\lambda t}\delta R_{2}(u-ct), \\
& \delta V(u-ct,t)=e^{\lambda t}\delta V_{1}(u-ct)+e^{\bar\lambda t}\delta
V_{2}(u-ct),
\end{split}
\end{equation}
where subscripts $1$ and its complex conjugate $2$ are used to distinguish different functions of $u$.  $\Re(\lambda)$ is the growth
rate of perturbation. Then
\begin{equation}\label{dRdVlambdabar}
\begin{split}
& \delta \bar R(u-ct,t)=e^{\bar \lambda t}\delta \bar R_{1}(u-ct)+e^{\lambda t}\delta \bar R_{2}(u-ct), \\
& \delta \bar V(u-ct,t)=e^{\bar \lambda t}\delta \bar
V_{1}(u-ct)+e^{\lambda t}\delta \bar V_{2}(u-ct).
\end{split}
\end{equation}
A dynamics of general perturbations can be represented as superposition of solutions with different $\lambda$. Thus our goal is to find possible values of $\lambda$.
Switching to the frame moving with the Stokes' wave and substituting~\eqref{dRdVlambda} and~\eqref{dRdVlambdabar}
into~\eqref{Reqn4}-\eqref{Veqn4} linearized with respect to perturbations, we collect terms $\sim e^{\lambda t}$ and obtain:%$\propto$
\begin{align}
\lambda \delta R_{1} &=c(\delta R_{1})_u+\nonumber\\
&+ \I \left[\delta U_{1}  R_u +U (\delta R_{1})_u - \delta R_{1} U_u - R( \delta U_{1})_u \right],\nonumber\\
\lambda \delta \bar R_{2} &=c(\delta \bar R_{2})_u-\label{RVeqn4lina}\\
&-\I \left[\delta \bar U_{2}  \bar R_u +\bar U (\delta \bar R_{2})_u - \delta \bar R_{2} \bar U_u - \bar R( \delta \bar U_{2})_u \right], \nonumber\\
\lambda \delta V_{1} &=c(\delta V_{1})_u+ g\delta R_{1}+\nonumber\\
&+\I \left[ \delta U_{1} V_u + U
(\delta V_{1})_u - \delta R_{1}B_u- R(\delta B_{1})_u\right ], \nonumber\\
\lambda \delta \bar V_{2} &= c(\delta \bar V_{2})_u+ g\delta \bar R_{2}-\nonumber\\
&-\I \left[ \delta\bar U_{2} \bar V_u + \bar U (\delta\bar V_{2})_u - \delta \bar R_{2}\bar B_u-
\bar R(\delta \bar B_{2})_u\right ],\nonumber
\end{align}
where
\begin{align}
\delta U_{1}&=\hat P^-(\delta R_{1}\bar V+R\delta\bar V_{2}+ {\delta \bar R_{2}}V+\bar R\delta V_{1}), \nonumber\\
\delta \bar U_{2}&=\hat P^+(\delta \bar R_{2} V+\bar R\delta V_{1}+
{\delta  R_{1}}\bar V+R\delta \bar V_{2}),
\nonumber \\%\label{Udef4lina}\\
\quad \delta B_{1}&= \hat P^-(\delta V _{1}\bar V+V\delta\bar V_{2}), \nonumber   \\
\quad \delta\bar B_{2}&= \hat P^+(\delta\bar  V _{2} V+\bar V\delta
V_{1}).\nonumber
\end{align}
Here $\hat P^{+}(f(u))\equiv\frac{1}{2}(1-\I \hat H)f$ is the projector onto the class of functions analytic in the upper half-plane $\C^{+}$ of $w$.

Equations~\eqref{RVeqn4lina} together with the periodicity of $\delta
R_{1}, \delta \bar R_{2}, \delta V_{1}, \delta \bar V_{2}$  in $u$ form
the eigenvalue problem for the eigenvector
\begin{equation} \label{deltaRVeigen}
\vec x = (\delta R_{1}, \delta \bar R_{2}, \delta V_{1}, \delta \bar V_{2})^{T},
\end{equation}
where $T$ means transposition.

\textit{Numerical Method. } For Stokes' wave solution we were reconstructing $R$ and $V$ in equations \eqref{RVeqn4lina}
using high precision Stokes' waves available at~\cite{PadePolesList}. Eigenvalue problem, given by the equations~\eqref{RVeqn4lina}, was solved by application of
shift-invert method in combination with Arnoldi algorithm for largest magnitude eigenvalues,
specifically ARPACK-NG (available at~\cite{ARPACK-NG}) realization was used. Full description of the method can be found in~\cite{KLSD2023}. We represent each of $\delta R_{1}, \delta \bar R_{2}, \delta V_{1},
\delta \bar V_{2}$ by a truncated Fourier series of $N$ Fourier harmonics.
Arnoldi algorithm is the most efficient, when it tries to locate few eigenvalues of largest magnitude. Let us suppose that we
have a guess of an eigenvalue $\sigma$. Then we can consider the modified eigenvalue problem~\cite{saad1992numerical}:
\begin{equation}
(A-\sigma I)^{-1}\vec x = \nu \vec x,\label{Arnoldi_modified}
\end{equation}
eigenvalues of which $\nu_j$ are related to the eigenvalues of original problem $\lambda_j$ by a simple formula:
\begin{equation}
\nu_j = \frac{1}{\lambda_j-\sigma}.
\end{equation}
It is clear, that if our guess $\sigma$ is close enough to the eigenvalue $\lambda_j$ we are looking for,
the magnitude of the $\nu_j$ eigenvalue will be the largest one.

Calculations would be unfeasible without application of inhomogeneous grid approach, which was originally introduced in~\cite{LDS2017}. It is also described in one of the appendices of~\cite{KLSD2023} more specifically for our formulation of the problem. Just as an illustration, resolution of $N=45000$ harmonics would correspond to $N\sim 10^{9}$ in the case of evenly spaced points. For convenience gravity acceleration $g=1$ in all computations.

\textit{Main Results. }
Taking into account scaling~\eqref{large_gr_scaling} and the value of the growth rate for the first branch of instability $\lambda_{1,s_{last}}\simeq 8.99799$ for the steepest wave reported in~\cite{KLSD2023} with steepness $s_{last}=H/\Lambda=0.1410597062844507$, the closest one to the threshold of the fourth branch of instability $s_4=0.141062741$, one would expect the value of the growth rate for the first (the strongest) branch of instability $\lambda_{1,s_4}\approx \lambda_{1,s_{last}} \sqrt{(s_{max}-s_{last})(s_{max}-s_4)}\simeq 20$. In the Stokes' wave collection~\cite{PadePolesList} only two waves with steepness beyond $s_4$ are available: with $s_{new,1}=0.1410627532159434$ and $s_{new,2}=0.1410629773898305$. Corresponding estimation of the growth rates for the first branch of superharmonic instability are
$\lambda_{1,s_{new,1}} \simeq 20.458$ and $\lambda_{1,s_{new,2}} \simeq 24.571$. As a result we have chosen a largest eigenvalues guess $\sigma=27.0$ for shift-invert algorithm. The limit on memory (RAM) on available computers was approximately 2 TiB = $2^{41}$ Bytes which resulted in close to maximum available value of number of harmonics on inhomogeneous grid $N=180224$ (we had to leave some reserve for operational system etc.). In order to check the stability of the obtained result with respect to resolution and also to estimate how many digits of accuracy we have, we used slightly smaller number of harmonics $N=163840$ as a test. Solution of the eigenvalue problem~\eqref{Arnoldi_modified} using Arnoldi algorithm to find 64  eigenvalues with largest in magnitudes was usually taking
approximately 40 hours on a computational workstation with 48 cores ($2\times$ AMD EPYC\texttrademark 74F3 at 3.2GHz). \red{We would like to note that although the number of harmonics 4 times larger than for the steepest wave in the previous paper~\cite{KLSD2023} (where the maximum amount of used memory was about 128 GiB), because we reconstruct the full matrix of linearized system~\eqref{RVeqn4lina}, the computation requires $16=4^2$ times more memory (2 TiB = 2048 GiB = $16\times128$ GiB).} 

For both available Stokes' waves the residue of the equations~\eqref{Reqn4}-\eqref{Veqn4} was significantly higher than in previous work~\cite{KLSD2023}, on the level of $10^{-8}$ due to the limit on memory, resulting in limit on number of harmonics. Better accuracy is achievable using more advanced approach~\cite{DS2023} not requiring creation of the whole matrix for the shift-invert method, which would allow to increase $N$ dramatically. So current Letter can be considered as the first approach to the problem. Results for both waves $s_{new,1}$ and $s_{new,2}$ are presented in Table 1 and Table 2. The rule for selection of the eigenvalues was the requirement of real part to be at least few order of magnitude larger, than the imaginary one. Taking into account relatively high residue for the original equation we were expecting to have relatively large nonzero imaginary parts. Additional way to check the correctness of the choice was to observe whether the real part stays practically constant with change of resolution, which was always confirmed as one can see.

\begin{table}
\center
\caption{Table 1. Instabilities growth rates for Stokes' wave with steepness $H/\Lambda=s_{new,1} = 0.1410627532159434$. Left column is an instability branch number. Upper values in every row are for lower resolution $N=163840$, lower ones are for higher resolution $N=180224$. The rightmost column shows the relative difference of the growth rate $\lambda$ between these two resolutions.}
\label{tab:0.1410627532159434}
\begin{tabular}{|c|c|c|c|}
\hline
\# & $\Re(\lambda)$ & $\Im(\lambda)$ & $|\Delta\lambda|/|\lambda|$ \\
\hline
1 & \begin{tabular}{l}
                     $20.5179856806$\\
                     $20.5179856816$\\
    \end{tabular} &  \begin{tabular}{l}
                     $3.2\times 10^{-10}$\\
                     $6.0\times 10^{-11}$\\
                     \end{tabular} &  $5\times 10^{-11}$\\
\hline
2 & \begin{tabular}{l}
                     $4.725163039323$\\
                     $4.725163039336$\\
    \end{tabular} &  \begin{tabular}{l}
                     $-3.0\times 10^{-11}$\\
                     $-6.5\times 10^{-11}$\\
                     \end{tabular} &  $3\times 10^{-12}$\\
\hline
3 & \begin{tabular}{l}
                     $1.0783329973$\\
                     $1.0783329964$\\
    \end{tabular} &  \begin{tabular}{l}
                     $-8.5\times 10^{-9}$\\
                     $-8.4\times 10^{-9}$\\
                     \end{tabular} &  $9\times 10^{-10}$\\
\hline
4 & \begin{tabular}{l}
                     $0.0418722$\\
                     $0.0418706$\\
    \end{tabular} &  \begin{tabular}{l}
                     $-1.3\times 10^{-5}$\\
                     $-1.4\times 10^{-5}$\\
                     \end{tabular} &  $4\times 10^{-5}$\\
\hline
\end{tabular}
\end{table}

In the case of a wave with steepness $H/\Lambda=s_{new,1} = 0.1410627532159434$, as calculated above, the approximation of the largest growth rate, corresponding to the first branch of superharmonic instability, using \eqref{large_gr_scaling} gives $\lambda_{1,s_{new,1}} \simeq 20.46$, while real value in Table 1 is $\approx 20.52$, thus relative difference is around $0.3\%$ which looks like very good accuracy for a simple asymptotic.

The approximation of the second branch of instability growth rate can be obtained as follows. For the steepest wave $s_{last}=H/\Lambda=0.1410597062844507$ reported in~\cite{KLSD2023} the growth rate for the second branch $\lambda_{2,s_{last}} \simeq 2.0678519897$, thus, using~\eqref{large_gr_scaling}, one can approximate $\lambda_{2,s_{new,1}}\approx \lambda_{2,s_{last}} \sqrt{(s_{max}-s_{last})(s_{max}-s_{new,1})}\simeq 4.7015$. Comparing with real value $4.7251630393$ in Table 1 we can state that relative accuracy of approximation is about $0.3\%$, like for the previous branch.

The reported in~\cite{KLSD2023} growth rate for the third branch of instability for the Stokes' wave of steepness $s_{last}$ was $\lambda_{3,s_{last}} \simeq 0.439824025$. It is not completely clear which formula to use, as we are right on the threshold of applicability between formulae~\eqref{small_gr_scaling} and~\eqref{large_gr_scaling}. Let us try both of them. Using~\eqref{large_gr_scaling} one gets $\lambda_{3,s_{new,1}}\approx \lambda_{3,s_{last}} \sqrt{(s_{max}-s_{last})(s_{max}-s_{new,1})}\simeq 0.999997$ which differs from the real value  $1.07833300$ in Table 1 by $7\%$. Using~\eqref{small_gr_scaling} for $x=(s_{max}-s_{new,1})/(s_{max}-s_3)$ one can get approximation $\lambda_{3,s_{new,1}}\approx 0.6375$ which is off by $41\%$. So one can conclude that~\eqref{small_gr_scaling} stops working well already before the threshold $\lambda<1$, while asymptotics~\eqref{large_gr_scaling} starts to work reasonably well already for $\lambda>0.5$.

And finally, for the new fourth branch of superharmonic instability we can use only formula~\eqref{small_gr_scaling}. Using value $x=(s_{max}-s_{new,1})/(s_{max}-s_4)$ one gets the approximation $\lambda_{4,s_{new,1}}\approx 0.04368$ which is close to $4\%$ off the real value $0.04187$ in Table 1.

\begin{table}
\center
\caption{Table 2. Instabilities growth rates for Stokes' wave with steepness $H/\Lambda=s_{new,2} = 0.1410629773898305$. As before, left column is an instability branch number; upper values in every row are for lower resolution $N=163840$, lower ones are for higher resolution $N=180224$. The rightmost column shows the relative difference of the growth rate $\lambda$ between these two resolutions.}
\label{tab:0.1410629773898305}
\begin{tabular}{|c|c|c|c|}
\hline
\# & $\Re(\lambda)$ & $\Im(\lambda)$ & $|\Delta\lambda|/|\lambda|$ \\
\hline
1 & \begin{tabular}{l}
                     $24.6467967$\\
                     $24.6467960$\\
    \end{tabular} &  \begin{tabular}{l}
                     $5.0\times 10^{-10}$\\
                     $2.2\times 10^{-10}$\\
                     \end{tabular} &  $3\times 10^{-8}$\\
\hline
2 & \begin{tabular}{l}
                     $5.67667745$\\
                     $5.67667730$\\
    \end{tabular} &  \begin{tabular}{l}
                     $-7.4\times 10^{-11}$\\
                     $-2.2\times 10^{-10}$\\
                     \end{tabular} &  $3\times 10^{-8}$\\
\hline
3 & \begin{tabular}{l}
                     $1.300748296$\\
                     $1.300748249$\\
    \end{tabular} &  \begin{tabular}{l}
                     $-6.1\times 10^{-9}$\\
                     $2.2\times 10^{-9}$\\
                     \end{tabular} &  $4\times 10^{-8}$\\
\hline
4 & \begin{tabular}{l}
                     $0.207468119$\\
                     $0.207468162$\\
    \end{tabular} &  \begin{tabular}{l}
                     $1.9\times 10^{-6}$\\
                     $2.3\times 10^{-6}$\\
                     \end{tabular} &  $2\times 10^{-7}$\\
\hline
\end{tabular}
\end{table}

For the wave with steepness $H/\Lambda=s_{new,2} = 0.1410629773898305$, which is significantly further from the threshold of the fourth branch of superharmonic instability $s_4$, the approximation of the largest growth rate, corresponding to the first branch of superharmonic instability, using \eqref{large_gr_scaling} (see full analysis above) gives $\lambda_{1,s_{new,2}} \simeq 24.57$, while real value in Table 2 is $\approx 24.65$, resulting in relative difference around $0.3\%$ (the same as for $s_{new,1}$).

The same analysis for the second branch of instability growth rate, using~\eqref{large_gr_scaling}, gives  $\lambda_{2,s_{new,2}}\approx \lambda_{2,s_{last}} \sqrt{(s_{max}-s_{last})(s_{max}-s_{new,2})}\simeq 5.647$, while real value in Table 2 is $5.676677$, which is only $0.5\%$ off.

Taking into account results for $\lambda_{3,s_{new,1}}$, namely that asymptotic~\eqref{large_gr_scaling} works \red{far} better than~\eqref{small_gr_scaling} even for smaller steepness $s_{new,1}$, it is reasonable to use it again. Using~\eqref{large_gr_scaling} one gets $\lambda_{3,s_{new,2}}\approx \lambda_{3,s_{last}} \sqrt{(s_{max}-s_{last})(s_{max}-s_{new,2})}\simeq 1.20$ which differs from the real value  $1.300748$ in Table 2 by less than $8\%$. If we use result for $s_{new,1}$ as a starting point, then approximation is way better: $\lambda_{3,s_{new,2}}\approx \lambda_{3,s_{new,1}} \sqrt{(s_{max}-s_{new,1})(s_{max}-s_{new,2})}\simeq 1.295$ which is only $0.4\%$ off. Thus, the formula~\eqref{large_gr_scaling} works very well already around the threshold $\lambda \approx 1$.

This time for the new fourth branch of superharmonic instability we can again use formula~\eqref{small_gr_scaling} with $x=(s_{max}-s_{new,2})/(s_{max}-s_4)$, which results in approximation $\lambda_{4,s_{new,2}}\approx 0.2149$. Comparing with the real value $0.207468$ in Table 2 we get accuracy $4\%$. While if we try to use scaling~\eqref{large_gr_scaling} and result for $\lambda_{4,s_{new,1}}$ from Table 1, we get very bad approximation $0.04187\sqrt{(s_{max}-s_{new,1})(s_{max}-s_{new,2})}\simeq 0.050$, corresponding to the fact that asymptotics~\eqref{large_gr_scaling} starts to work well only for $\lambda \approx 1$ or at least $\lambda >  0.5$.

To summarize, we have reached for the first time the fourth branch of superharmonic instability of the Stokes' wave. Taking into account limited computational resources, this work can be considered as the first approach to the problem. The relative accuracy of the obtained new branch of instability growth rate is of the order of $10^{-5}$ and $10^{-7}$ for two available Stokes' waves beyond the threshold of instability (relative difference in steepness from the limiting Stokes' wave for the steeper wave $s_{new,2}$ was less than $3.59\times 10^{-6}$). This is relatively low accuracy in comparison with previous work~\cite{KLSD2023}, which considered only the first three branches of instability, thus requiring approximately 4 times less harmonics.
Nevertheless, we were able to check both phenomenological relations approximating growth rates, which were proposed in~\cite{KLSD2023}, namely~\eqref{large_gr_scaling} for $\lambda > 1$ and~\eqref{small_gr_scaling} for $\lambda < 1$. It appears that asymptotics~\eqref{large_gr_scaling} works acceptably already for $\lambda > 0.5$ (accuracy better than $10\%$) and very well for $\lambda > 1$ (accuracy better than $0.4\%$), while expression~\eqref{small_gr_scaling} gives very reasonable approximation of the instability growth rate at the inception of a new branch of superharmonic instability of the Stokes' wave ($\lambda < 0.5$). \red{It means that there is no need to perform similar computations for further (infinitely many) branches of instability, in the cases when accuracy better than 10\% is acceptable. After the inception of instability branch, one can use~\eqref{small_gr_scaling} up to growth rate $\lambda=0.5$ and continue for $\lambda>0.5$ using asymptotics~\eqref{large_gr_scaling}.}

As a next step authors consider implementation of more advanced and memory efficient algorithms, similar to~\cite{DS2023}, which would allow to use significantly better resolution, resulting in much lower errors in representation of the unperturbed Stokes' wave. Another necessary step for accuracy improvement is to improve accuracy of the positions of the Hamiltonian's extrema, corresponding to appearance of the new branches of instability. Currently available values, reported in~\cite{KLSD2023}, are given with an accuracy of (at least) around 7 digits after decimal point, while already in this paper we considered Stokes' waves which differ from the limiting Stokes' wave only 6th digit, which might influence accuracy asymptotics~\eqref{small_gr_scaling} analysis.  

We acknowledge the computing time provided to us at computer facilities
at Landau Institute. Authors would like to thank the Russian Science Foundation for funding this research in the framework of the grant 25-72-31023. Also authors would like to thank developers of FFTW~\cite{FFTW}, ARPACK-NG~\cite{ARPACK-NG}, and the whole GNU project~\cite{GNU} for developing, and supporting this useful and free software.

%\bibliographystyle{aipnum4-2}
%\bibliographystyle{apsrev4-2}
%\bibliography{KAO_complete_list_of_papers,surfacewaves}

\begin{thebibliography}{19}%
\makeatletter
\providecommand \@ifxundefined [1]{%
 \@ifx{#1\undefined}
}%
\providecommand \@ifnum [1]{%
 \ifnum #1\expandafter \@firstoftwo
 \else \expandafter \@secondoftwo
 \fi
}%
\providecommand \@ifx [1]{%
 \ifx #1\expandafter \@firstoftwo
 \else \expandafter \@secondoftwo
 \fi
}%
\providecommand \natexlab [1]{#1}%
\providecommand \enquote  [1]{``#1''}%
\providecommand \bibnamefont  [1]{#1}%
\providecommand \bibfnamefont [1]{#1}%
\providecommand \citenamefont [1]{#1}%
\providecommand \href@noop [0]{\@secondoftwo}%
\providecommand \href [0]{\begingroup \@sanitize@url \@href}%
\providecommand \@href[1]{\@@startlink{#1}\@@href}%
\providecommand \@@href[1]{\endgroup#1\@@endlink}%
\providecommand \@sanitize@url [0]{\catcode `\\12\catcode `\$12\catcode
  `\&12\catcode `\#12\catcode `\^12\catcode `\_12\catcode `\%12\relax}%
\providecommand \@@startlink[1]{}%
\providecommand \@@endlink[0]{}%
\providecommand \url  [0]{\begingroup\@sanitize@url \@url }%
\providecommand \@url [1]{\endgroup\@href {#1}{\urlprefix }}%
\providecommand \urlprefix  [0]{URL }%
\providecommand \Eprint [0]{\href }%
\providecommand \doibase [0]{https://doi.org/}%
\providecommand \selectlanguage [0]{\@gobble}%
\providecommand \bibinfo  [0]{\@secondoftwo}%
\providecommand \bibfield  [0]{\@secondoftwo}%
\providecommand \translation [1]{[#1]}%
\providecommand \BibitemOpen [0]{}%
\providecommand \bibitemStop [0]{}%
\providecommand \bibitemNoStop [0]{.\EOS\space}%
\providecommand \EOS [0]{\spacefactor3000\relax}%
\providecommand \BibitemShut  [1]{\csname bibitem#1\endcsname}%
\let\auto@bib@innerbib\@empty
%</preamble>
\bibitem [{\citenamefont {Stokes}(1847)}]{Stokes1847}%
  \BibitemOpen
  \bibfield  {author} {\bibinfo {author} {\bibfnamefont {G.~G.}\ \bibnamefont
  {Stokes}},\ }\href@noop {} {\bibfield  {journal} {\bibinfo  {journal}
  {Transactions of the Cambridge Philosophical Society}\ }\textbf {\bibinfo
  {volume} {8}},\ \bibinfo {pages} {441} (\bibinfo {year} {1847})}\BibitemShut
  {NoStop}%
\bibitem [{\citenamefont {Stokes}(1880)}]{Stokes1880}%
  \BibitemOpen
  \bibfield  {author} {\bibinfo {author} {\bibfnamefont {G.~G.}\ \bibnamefont
  {Stokes}},\ }\href@noop {} {\bibfield  {journal} {\bibinfo  {journal}
  {Mathematical and Physical Papers}\ }\textbf {\bibinfo {volume} {1}},\
  \bibinfo {pages} {197} (\bibinfo {year} {1880})}\BibitemShut {NoStop}%
\bibitem [{\citenamefont {Longuet-Higgins}\ and\ \citenamefont
  {Selwyn}(1978)}]{LHS1978-1}%
  \BibitemOpen
  \bibfield  {author} {\bibinfo {author} {\bibnamefont {Longuet-Higgins}}\ and\
  \bibinfo {author} {\bibfnamefont {M.}~\bibnamefont {Selwyn}},\ }\href@noop {}
  {\bibfield  {journal} {\bibinfo  {journal} {Proceedings of the Royal Society
  of London. A. Mathematical and Physical Sciences}\ }\textbf {\bibinfo
  {volume} {360}},\ \bibinfo {pages} {471} (\bibinfo {year}
  {1978})}\BibitemShut {NoStop}%
\bibitem [{\citenamefont {Tanaka}(1983)}]{Tanaka1983}%
  \BibitemOpen
  \bibfield  {author} {\bibinfo {author} {\bibfnamefont {M.}~\bibnamefont
  {Tanaka}},\ }\href@noop {} {\bibfield  {journal} {\bibinfo  {journal}
  {Journal of the physical society of Japan}\ }\textbf {\bibinfo {volume}
  {52}},\ \bibinfo {pages} {3047} (\bibinfo {year} {1983})}\BibitemShut
  {NoStop}%
\bibitem [{\citenamefont {Saffman}(1985)}]{Saffman1985}%
  \BibitemOpen
  \bibfield  {author} {\bibinfo {author} {\bibfnamefont {P.}~\bibnamefont
  {Saffman}},\ }\href@noop {} {\bibfield  {journal} {\bibinfo  {journal}
  {{Journal of Fluid Mechanics}}\ }\textbf {\bibinfo {volume} {159}},\ \bibinfo
  {pages} {169} (\bibinfo {year} {1985})}\BibitemShut {NoStop}%
\bibitem [{\citenamefont {Longuet-Higgins}\ and\ \citenamefont
  {Tanaka}(1997)}]{LHT1997}%
  \BibitemOpen
  \bibfield  {author} {\bibinfo {author} {\bibfnamefont {M.}~\bibnamefont
  {Longuet-Higgins}}\ and\ \bibinfo {author} {\bibfnamefont {M.}~\bibnamefont
  {Tanaka}},\ }\href@noop {} {\bibfield  {journal} {\bibinfo  {journal}
  {{J.\,Fluid\,Mech.}}\ }\textbf {\bibinfo {volume} {336}},\ \bibinfo {pages}
  {51} (\bibinfo {year} {1997})}\BibitemShut {NoStop}%
\bibitem [{\citenamefont {Korotkevich}\ \emph {et~al.}(2023)\citenamefont
  {Korotkevich}, \citenamefont {Lushnikov}, \citenamefont {Semenova},\ and\
  \citenamefont {Dyachenko}}]{KLSD2023}%
  \BibitemOpen
  \bibfield  {author} {\bibinfo {author} {\bibfnamefont {A.~O.}\ \bibnamefont
  {Korotkevich}}, \bibinfo {author} {\bibfnamefont {P.~M.}\ \bibnamefont
  {Lushnikov}}, \bibinfo {author} {\bibfnamefont {A.}~\bibnamefont
  {Semenova}},\ and\ \bibinfo {author} {\bibfnamefont {S.~A.}\ \bibnamefont
  {Dyachenko}},\ }\href {https://doi.org/10.1111/sapm.12535} {\bibfield
  {journal} {\bibinfo  {journal} {Studies in Applied Mathematics}\ }\textbf
  {\bibinfo {volume} {150}},\ \bibinfo {pages} {119} (\bibinfo {year}
  {2023})},\ \Eprint
  {https://arxiv.org/abs/https://onlinelibrary.wiley.com/doi/pdf/10.1111/sapm.12535}
  {https://onlinelibrary.wiley.com/doi/pdf/10.1111/sapm.12535} \BibitemShut
  {NoStop}%
\bibitem [{\citenamefont {Deconinck}\ \emph {et~al.}(2023)\citenamefont
  {Deconinck}, \citenamefont {Dyachenko}, \citenamefont {Lushnikov},\ and\
  \citenamefont {Semenova}}]{PNAS2023}%
  \BibitemOpen
  \bibfield  {author} {\bibinfo {author} {\bibfnamefont {B.}~\bibnamefont
  {Deconinck}}, \bibinfo {author} {\bibfnamefont {S.~A.}\ \bibnamefont
  {Dyachenko}}, \bibinfo {author} {\bibfnamefont {P.~M.}\ \bibnamefont
  {Lushnikov}},\ and\ \bibinfo {author} {\bibfnamefont {A.}~\bibnamefont
  {Semenova}},\ }\href {https://doi.org/10.1073/pnas.2308935120} {\bibfield
  {journal} {\bibinfo  {journal} {Proceedings of the National Academy of
  Sciences}\ }\textbf {\bibinfo {volume} {120}},\ \bibinfo {pages}
  {e2308935120} (\bibinfo {year} {2023})}\BibitemShut {NoStop}%
\bibitem [{\citenamefont {Lushnikov}\ \emph {et~al.}(2017)\citenamefont
  {Lushnikov}, \citenamefont {Dyachenko},\ and\ \citenamefont
  {A.~Silantyev}}]{LDS2017}%
  \BibitemOpen
  \bibfield  {author} {\bibinfo {author} {\bibfnamefont {P.~M.}\ \bibnamefont
  {Lushnikov}}, \bibinfo {author} {\bibfnamefont {S.~A.}\ \bibnamefont
  {Dyachenko}},\ and\ \bibinfo {author} {\bibfnamefont {D.}~\bibnamefont
  {A.~Silantyev}},\ }\href@noop {} {\bibfield  {journal} {\bibinfo  {journal}
  {Proceedings of the Royal Society A: Mathematical, Physical and Engineering
  Sciences}\ }\textbf {\bibinfo {volume} {473}},\ \bibinfo {pages} {20170198}
  (\bibinfo {year} {2017})}\BibitemShut {NoStop}%
\bibitem [{\citenamefont {Dyachenko}\ \emph {et~al.}(2016)\citenamefont
  {Dyachenko}, \citenamefont {Lushnikov},\ and\ \citenamefont
  {Korotkevich}}]{DLK2016}%
  \BibitemOpen
  \bibfield  {author} {\bibinfo {author} {\bibfnamefont {S.~A.}\ \bibnamefont
  {Dyachenko}}, \bibinfo {author} {\bibfnamefont {P.~M.}\ \bibnamefont
  {Lushnikov}},\ and\ \bibinfo {author} {\bibfnamefont {A.~O.}\ \bibnamefont
  {Korotkevich}},\ }\href {https://doi.org/10.1111/sapm.12128} {\bibfield
  {journal} {\bibinfo  {journal} {Studies in Applied Mathematics}\ }\textbf
  {\bibinfo {volume} {137}},\ \bibinfo {pages} {419} (\bibinfo {year}
  {2016})},\ \Eprint
  {https://arxiv.org/abs/https://onlinelibrary.wiley.com/doi/pdf/10.1111/sapm.12128}
  {https://onlinelibrary.wiley.com/doi/pdf/10.1111/sapm.12128} \BibitemShut
  {NoStop}%
\bibitem [{\citenamefont {Dyachenko}\ \emph {et~al.}(2025)\citenamefont
  {Dyachenko}, \citenamefont {Korotkevich}, \citenamefont {Lushnikov},
  \citenamefont {Semenova},\ and\ \citenamefont {Silantyev}}]{PadePolesList}%
  \BibitemOpen
  \bibfield  {author} {\bibinfo {author} {\bibfnamefont {S.~A.}\ \bibnamefont
  {Dyachenko}}, \bibinfo {author} {\bibfnamefont {A.~O.}\ \bibnamefont
  {Korotkevich}}, \bibinfo {author} {\bibfnamefont {P.~M.}\ \bibnamefont
  {Lushnikov}}, \bibinfo {author} {\bibfnamefont {A.}~\bibnamefont
  {Semenova}},\ and\ \bibinfo {author} {\bibfnamefont {D.}~\bibnamefont
  {Silantyev}},\ }\href {http://stokeswave.org} {\bibfield  {journal} {\bibinfo
   {journal} {{\tt http://stokeswave.org}}\ } (\bibinfo {year}
  {2015-2025})}\BibitemShut {NoStop}%
\bibitem [{\citenamefont {Dyachenko}\ \emph {et~al.}(2013)\citenamefont
  {Dyachenko}, \citenamefont {Lushnikov},\ and\ \citenamefont
  {Korotkevich}}]{DLK2013}%
  \BibitemOpen
  \bibfield  {author} {\bibinfo {author} {\bibfnamefont {S.~A.}\ \bibnamefont
  {Dyachenko}}, \bibinfo {author} {\bibfnamefont {P.~M.}\ \bibnamefont
  {Lushnikov}},\ and\ \bibinfo {author} {\bibfnamefont {A.~O.}\ \bibnamefont
  {Korotkevich}},\ }\href {https://doi.org/10.1134/S0021364013240077}
  {\bibfield  {journal} {\bibinfo  {journal} {JETP Letters}\ }\textbf {\bibinfo
  {volume} {98}},\ \bibinfo {pages} {675} (\bibinfo {year} {2013})}\BibitemShut
  {NoStop}%
\bibitem [{\citenamefont {Tanveer}(1991)}]{Tanveer1991}%
  \BibitemOpen
  \bibfield  {author} {\bibinfo {author} {\bibfnamefont {S.}~\bibnamefont
  {Tanveer}},\ }\href@noop {} {\bibfield  {journal} {\bibinfo  {journal} {Proc.
  R. Soc. Lond. A}\ }\textbf {\bibinfo {volume} {435}},\ \bibinfo {pages} {137}
  (\bibinfo {year} {1991})}\BibitemShut {NoStop}%
\bibitem [{\citenamefont {Dyachenko}(2001)}]{Dyachenko2001}%
  \BibitemOpen
  \bibfield  {author} {\bibinfo {author} {\bibfnamefont {A.~I.}\ \bibnamefont
  {Dyachenko}},\ }\href@noop {} {\bibfield  {journal} {\bibinfo  {journal}
  {Dokl. Math.}\ }\textbf {\bibinfo {volume} {63}},\ \bibinfo {pages} {115}
  (\bibinfo {year} {2001})}\BibitemShut {NoStop}%
\bibitem [{\citenamefont {ARPACK-NG}(2025)}]{ARPACK-NG}%
  \BibitemOpen
  \bibfield  {author} {\bibinfo {author} {\bibnamefont {ARPACK-NG}},\ }\href
  {https://github.com/opencollab/arpack-ng} {\bibfield  {journal} {\bibinfo
  {journal} {github.com}\ } (\bibinfo {year} {2025})},\  {{\tt
  https://github.com/opencollab/arpack-ng}} \BibitemShut {NoStop}%
\bibitem [{\citenamefont {Saad}(1992)}]{saad1992numerical}%
  \BibitemOpen
  \bibfield  {author} {\bibinfo {author} {\bibfnamefont {Y.}~\bibnamefont
  {Saad}},\ }\href@noop {} {\emph {\bibinfo {title} {{Numerical methods for
  large eigenvalue problems}}}}\ (\bibinfo  {publisher} {Manchester University
  Press},\ \bibinfo {year} {1992})\BibitemShut {NoStop}%
\bibitem [{\citenamefont {Dyachenko}\ and\ \citenamefont
  {Semenova}(2023)}]{DS2023}%
  \BibitemOpen
  \bibfield  {author} {\bibinfo {author} {\bibfnamefont {S.~A.}\ \bibnamefont
  {Dyachenko}}\ and\ \bibinfo {author} {\bibfnamefont {A.}~\bibnamefont
  {Semenova}},\ }\href {https://doi.org/https://doi.org/10.1111/sapm.12554}
  {\bibfield  {journal} {\bibinfo  {journal} {Studies in Applied Mathematics}\
  }\textbf {\bibinfo {volume} {150}},\ \bibinfo {pages} {705} (\bibinfo {year}
  {2023})}\BibitemShut {NoStop}%
\bibitem [{\citenamefont {Frigo}\ and\ \citenamefont {Johnson}(2005)}]{FFTW}%
  \BibitemOpen
  \bibfield  {author} {\bibinfo {author} {\bibfnamefont {M.}~\bibnamefont
  {Frigo}}\ and\ \bibinfo {author} {\bibfnamefont {S.~G.}\ \bibnamefont
  {Johnson}},\ }\href {http://fftw.org} {\bibfield  {journal} {\bibinfo
  {journal} {Proc.\,IEEE}\ }\textbf {\bibinfo {volume} {93}},\ \bibinfo {pages}
  {216} (\bibinfo {year} {2005})}\BibitemShut {NoStop}%
\bibitem [{\citenamefont {{GNU Project}}(2025)}]{GNU}%
  \BibitemOpen
  \bibfield  {author} {\bibinfo {author} {\bibnamefont {{GNU Project}}},\
  }\href {http://gnu.org} {\bibfield  {journal} {\bibinfo  {journal} {{\tt
  http://gnu.org}}\ } (\bibinfo {year} {1984-2025})}\BibitemShut {NoStop}%
\end{thebibliography}

%

\end{document}